\documentstyle[12pt,twoside,graphicx]{article}
\begin{document}

\begin{flushright}
DF/UFES-P005/97\\
gr-qc/9711047
\end{flushright}

\begin{center}

{\Large\bf SINGULARITIES AND CLASSICAL LIMIT\\[5pt]
IN QUANTUM COSMOLOGY\\[5pt]
WITH SCALAR FIELDS}
\medskip

{\bf Roberto Colistete Jr.\footnote{e-mail: roberto@cce.ufes.br}, J\'ulio 
C. Fabris\footnote{e-mail: fabris@cce.ufes.br}}
\medskip

Departamento de F\'{\i}sica, Universidade Federal do Esp\'{\i}rito Santo, 
29060-900, Vit\'oria, Esp\'{\i}rito Santo, Brazil
\medskip

and\\
{\bf Nelson Pinto-Neto\footnote{e-mail: nen@lca1.drp.cbpf.br}}
\medskip

Lafex, Centro Brasileiro de Pesquisas F\'{\i}sicas, Rua Xavier Sigaud 150,
22290-180, Rio de Janeiro, Brazil

\end{center}

\begin{abstract}
Minisuperspace models derived from Kaluza-Klein theories and low
energy string theory are studied. They are equivalent to one and two
minimally
coupled scalar fields. The general classical and quantum
solutions are obtained. 
Gaussian superposition of WKB solutions are constructed. Contrarily to
what is usually expected, these states are sharply peaked around the
classical
trajectories only for small values of the scale factor. This behaviour is
confirmed in the framework of the causal interpretation:
the Bohmian trajectories of many quantum states are
classical for small values of the scale factor but present quantum
behaviour when the scale factor becomes large. A consequence of this fact
is that these states present an initial singularity. However, there are
some particular superpositions of these wave functions which
have Bohmian trajectories without singularities.  There 
are also singular Bohmian trajectories with a short period 
of inflation which grow
forever. We could not find any non-singular trajectory which grows to the 
size of our universe.

\vspace{0.7cm}
PACS number(s): 04.20.Cv., 04.20.Me
\end{abstract}

\section{Introduction}

One of the main motivations to study quantum cosmology is to
investigate
if quantum gravitational effects can avoid the singularities which are
present in classical cosmological models \cite{QC}. If this is indeed the
case for the
initial
singularity, the next step should be to find in what conditions the
universe recovers its classical behaviour, yielding the large classical
expanding
universe we live in. In this paper we investigate these problems in the
framework of minisuperspace models with scalar fields as sources of the
gravitational field.

As a first example, we took a non-massive, minimally coupled scalar field,
in a
Friedman-Robertson-Walker universe with spacelike sections with positive
constant curvature. This model can be viewed as an effective
multidimensional
theory where the scalar field is understood as the scale factor of
internal
dimensions  \cite{multi}, or as a Brans-Dicke model redefined by a
conformal transformation \cite{brans}. We were
able to find the general classical solutions. All of them present initial
and final singularities. The model is quantized in the Dirac
way,
with arbitrary factor ordering, and the general solution of the
corresponding
Wheeler-DeWitt equation is found. To interpret the solutions, we first
adopted the `peak interpretation', where a prediction is made when the
wave function
is sharply peaked in a region and almost zero outside this region
\cite{peak}. A gaussian superposition of WKB solutions was constructed.
By employing the stationary phase condition, we were able to show that
this superposition
is sharply peaked around the classical trajectory only for small values of
the scale factor. Hence, contrarily to what is usually expected, the
classical
limit is recovered for small values of the scale factor, not for large
ones.
A consequence of this fact is that the initial classical singularities
continue
to be present at the quantum level. In order to confirm this strange
behaviour,
we also adopted an alternative interpretation of quantum mechanics
which was not constructed for cosmology but which can be easily applied
to a single system: it is the causal or the Bohm-de Broglie interpretation
of quantum mechanics \cite{boh}. It is completely different from the
others
because it is an ontological interpretation of quantum mechanics.
In the case of non-relativistic particles, the quantum particles follows 
a real trajectory,
independently of any observations, and it is accompanied by a wave
function.
The quantum effects are brought about by a quantum potential,
which can be derived from the Schr\"{o}dinger equation. It is a rather
simple interpretation which can be easily applied to minisuperspace
models \cite{bola1}. In this case, the Schr\"{o}dinger equation is
replaced
by the Wheeler-DeWitt equation, and the quantum trajectories are the time
evolutions of the metric and field variables, which obey a Hamilton-Jacobi
equation with an extra quantum potential term. The application of this
interpretation to some of the quantum solutions of our problem shows
exactly the same behaviour as found previously: the Bohmian trajectories
behave classically for small values of the scale factor while the quantum
behaviour appears when the scale factor becomes large. Singularities are
still present. However, when
we make superpositions of these wave functions, the initial singularity
disappears for some special cases, but none of these special trajectories
grows to the size of our universe.

The other case studied involves two minimally coupled scalar fields. They
can be viewed as a tree level effective action of string theory where
the second scalar field comes from the Kalb-Rammond
field \cite{cordas}. They can also be understood as generalized 
Brans-Dicke type models, which can be derived from compactification
of multidimensional theories with external gauge fields
\cite{brans2}. The results obtained in this case were
analogous to the preceding one. Along the lines of the peak
interpretation, gaussian WKB superpositions predicts a classical universe
for
small values of the scale factor
because they are peaked around the classical trajectories in this region.
Adopting the causal interpretation to investigate the singularity
problem, we found, as before, that many of the solutions present classical
behaviour when the scale factor is small (and hence singularities) but
behaves quantum mechanically when the scale factor becomes large.

This paper is organized as follows: in the next section we describe
the classical minisuperspace models of both one and two scalar fields
models, presenting their general classical solutions. In section 3
we quantize these models obtaining their corresponding Wheeler-DeWitt
equations and their respective general solutions. In section 4, the
gaussian
superpositions of WKB solutions are constructed and their peak along the
classical trajectories are exhibited. In section 5, the causal
interpretation
of quantum cosmology is shortly reviewed and applied to the quantum
solutions.
We end with comments and conclusions.
 
\section{The Classical Models}

Models with two scalar fields that interact non-trivially between
themselves
can be obtained from different theoretical contexts. Considering
Kaluza-Klein
supergravity theories, keeping just the bosonic sector, and reducing to
four dimensions, leads to effective actions with gravity plus two scalar
fields, one of them coupled non-minimally to the Einstein-Hilbert
lagrangian; the two scalar fields have an interaction between them.
More generally, every time we consider multidimensional models with
gauge fields, and reduce them to four dimensions, we find such
structure.
String theories, in particular, have an effective action in four
dimensions
given by the expression,
\begin{equation}
{\it L} = \sqrt{-g}e^{-\phi}\biggr(R + \phi_{;\rho}\phi^{;\rho} -
\frac{1}{12}H_{\mu\nu\lambda}H^{\mu\nu\lambda}\biggl)
\quad ,
\end{equation}
where $\phi$ is a dilaton field and $H_{\mu\nu\lambda}$ is a
Kalb-Ramond field which in four dimensions is equivalent to a scalar
field $\xi$
\par
In order to keep contact with this variety of models, all of them
having great importance in high energy conditions,
we will consider the general lagrangian
\begin{equation}
\label{lg}
{\it L} = \sqrt{-g}\biggr(\phi R - \omega\frac{\phi_{;\rho}\phi^{;\rho}}
{\phi} - \frac{\xi_{;\rho}\xi^{;\rho}}{\phi}\biggl) \quad ,
\end{equation}
where $\omega$ is a coupling constant. We remark the non-trivial
interaction between
$\phi$ and $\xi$. For the string effective
action, $\omega = - 1$ and for Kaluza-Klein theories $\omega =
\frac{1-d}{d}$,
where $d$ is the dimension of internal compact spacelike dimensions.
If we perform a conformal transformation such that
$g_{\mu\nu} = \phi^{-1}{\bar g}_{\mu\nu}$, we obtain the lagrangian
\begin{equation}
\label{lg2}
{\it L} = \sqrt{-g}\biggr[R - (\omega + \frac{3}{2})\frac{\phi_{;\rho}
\phi^{;\rho}}{\phi^2} - \frac{\xi_{;\rho}\xi^{;\rho}}{\phi^2}\biggl]
\quad ,
\end{equation}
where the bars have been suppressed.
From Eq. (\ref{lg2}) we deduce the field equations,
\begin{eqnarray}
R_{\mu\nu} - \frac{1}{2}g_{\mu\nu}R &=& \frac{\kappa}{\phi^2}
\biggr(\phi_{;\mu}\phi_{;\nu} - \frac{1}{2}g_{\mu\nu}\phi_{;\rho}
\phi^{;\rho}\biggl) + \frac{1}{\phi^2}\biggr(\xi_{;\mu}\xi_{;\nu}
- \frac{1}{2}g_{\mu\nu}\xi_{;\rho}\xi^{;\rho}\biggl) \quad ,\\
\Box\phi - \frac{\phi^{;\rho}\phi_{;\rho}}{\phi} +
\frac{\xi_{;\rho}\xi^{;\rho}}{\kappa\phi} &=& 0 \quad ,\\
\Box\xi - 2\frac{\xi_{;\rho}\phi^{;\rho}}{\phi} &=& 0 \quad ,
\end{eqnarray}
where $\kappa = \omega + \frac{3}{2}$.
\par
We consider now the Robertson-Walker metric
\begin{equation}
\label{m}
ds^2 = -N^2 {\rm d}t^2 + \frac{{a(t)}^2}{1 + \frac{\epsilon}{4}r^2}[{\rm
d}r^2 + r^2
({\rm d} \theta ^2 + \sin ^2 (\theta) {\rm d} \varphi ^2)],
\end{equation}
where the spatial curvature $\epsilon$ takes the values $0$, $1$,$-1$.
The equations of motion are, for $N=1$,
\begin{eqnarray}
\label{eq1}
3(\frac{\dot a}{a})^2 + \frac{3\epsilon}{a^2} &=&  
\frac{\kappa}{2}(\frac{\dot\phi}{\phi})^2 + 
\frac{1}{2}(\frac{\dot\xi}{\phi})^2 \quad ,\\
\label{eq2}
\ddot\phi + 3\frac{\dot a}{a}\dot\phi  - \frac{\dot\phi^2}{\phi} +
\frac{{\dot\xi}^2}{\kappa\phi} &=& 0
\quad ,\\
\label{eq3}
\ddot\xi + 3\frac{\dot a}{a}\dot\xi - 2\frac{\dot\phi}{\phi}\dot\xi &=& 0
\quad.
\end{eqnarray}
We will be interested in the case $\epsilon = 1$.
In what follows we will consider separately the cases where
$\xi = $ const. and $\xi \neq $ const..
\subsection{One scalar field minimaly coupled to gravity}
Henceforth, we consider in Eqs. (\ref{eq1},\ref{eq2},\ref{eq3}) $\xi = $
constant.
The solutions of the resulting equations can be easily found if we
reparametrize the time coordinate as $dt = a^3d\theta$. The integration
procedure is standard, and we just give the final results:
\begin{eqnarray}
\label{fi1}
\phi &=& Ae^{\theta + B} \quad ; \\
\label{cccp1}
a &=& \sqrt{A}\biggr(\frac{\kappa}{6}\biggl)^\frac{1}{4}
\frac{1}{\sqrt{\cosh\biggr(\sqrt{\frac{2}{3}}A\sqrt{\kappa}(\theta + 
C)\biggl)}} \quad .
\end{eqnarray}
In these expressions, $A$, $B$ and $C$ are integration constants. 
The universe expands from an initial singularity untill a maximum size and
then
contract to a final singularity. Note that
$a \propto t^\frac{1}{3}$ for small $a$. For $A=1$ and $B=C$, we obtain
the implicit relation:
\begin{equation}
\label{implicit1}
a(\phi)=\biggr[\frac{2}{3}\kappa\biggl]^\frac{1}{4}
\sqrt{\frac{\phi^{\sqrt{\frac{2}{3}\kappa}}}{1+\phi^{2\sqrt{\frac{2}{3}
\kappa}}}}\quad .
\end{equation}

\subsection{Two scalar fields coupled to gravity}
Considering the fields $\phi$ and $\xi$ in equations (\ref{eq1},
\ref{eq2},\ref{eq3}), and using again the same parameter $\theta$ as
defined previously, we find the following solutions:
\begin{eqnarray}
\label{flu1}
\xi &=& A + \frac{C}{B}\kappa\tanh\biggr(\frac{C(\theta +
D)}{\kappa}\biggl) \quad ;\\
\label{flu2}
\phi &=& \frac{C}{B}\frac{1}{\cosh\biggr(\frac{C(\theta +
D)}{\sqrt{\kappa}}
\biggl)} \quad ;\\
\label{flu3}
a &=& 
\frac{\sqrt{C}}{6^\frac{1}{4}}\frac{1}{\sqrt{\cosh\biggr(\sqrt{\frac{2}{3}}
|C|(\theta + E)\biggl)}} \quad .
\end{eqnarray}
In these expressions $A$, $B$, $C$, $D$ and $E$ are constants.
The qualitative behaviour of the scale factor is the same as in the
preceding
case (compare (\ref{cccp1}) with (\ref{flu3})). Again we have
$a \propto t^\frac{1}{3}$ when $a$ is small.
For $A = 0$, $B= C$, $D = E$ and $\kappa = \frac{3}{2}$, we can find
a simple implicit relation between $a$, $\phi$ and $\xi$:
\begin{eqnarray}
\label{che1}
\phi(a) &=& \frac{1}{|C|}\sqrt{6}a^2 \quad ;\\
\label{che2}
a(\xi) &=& \sqrt{|C|}(\frac{1}{6} - \frac{\xi^2}{9})^\frac{1}{4} \quad ;\\
\label{che3}
\phi(\xi) &=& \sqrt{1 - \frac{2}{3}\xi^2} \quad .
\end{eqnarray}
These implicit classical relations, together with Eq. (\ref{implicit1}),
will 
be compared with the trajectory on which the semi-classical wave function
of
the corresponding quantum model is peaked.

\section{Quantum Solutions in Minisuperspace}

We return to the lagrangian (\ref{lg2}) and we insert on it the
metric (\ref{m}).
The action takes the form
\begin{equation}
S = \int{Ldt}
\end{equation}
where
\begin{equation}
\label{lq}
L = \frac{12a{\dot a}^2}{N} - (3 + 
2\omega)\frac{a^3{\dot\phi}^2}{N\phi^2}
- 2\frac{a^3{\dot\xi}^2}{N\phi^2} - 12Na \quad .
\end{equation}
\par
From (\ref{lq}) we obtain the conjugate momenta,
\begin{eqnarray}
\pi_a &=& 24\frac{a\dot a}{N} \quad , \\
\pi_\phi &=& - 2(3+2\omega)\frac{a^3\dot\phi}{N\phi^2} \quad ,\\
\pi_\xi &=& -4\frac{a^3\dot\xi}{N\phi^2} \quad .
\end{eqnarray}
We can now construct the hamiltonian $H$, which takes the form
\begin{equation}
H = N\biggr[\frac{\pi_a^2}{48a} - \frac{\phi^2\pi_\phi^2}
{4(3+2\omega)a^3} - \frac{\phi^2\pi_\xi^2}{8a^3} +
12a\biggl] \equiv N{\cal H} \quad .
\end{equation}
Variation of $N$ yields the first class constraint ${\cal H} \approx 0$.
The Dirac quantization procedure yields the Wheeler-DeWitt equation  by 
imposing the condition:
\begin{equation}
\hat{{\cal H}} \Psi = 0 
\end{equation}
and performing the substitutions
\begin{eqnarray}
\pi_a^2 &\rightarrow& - \frac{\partial^2 }{\partial a^2} - \frac{p}{a}
\frac{\partial }{\partial a} \quad , \\
\pi_\phi^2 &\rightarrow& - \frac{\partial^2 }{\partial\phi^2}
- \frac{q}{\phi}\frac{\partial }{\partial\phi} \quad , \\
\pi_\xi^2 &\rightarrow& - \frac{\partial^2 }{\partial\xi^2}
\quad ,
\end{eqnarray}
where $p$ and $q$ are ordering factors. We have set $\hbar = 1$.
The Wheeler-DeWitt equation in the minisuperspace reads
\begin{equation}
\frac{a^2}{12}\biggr[\Psi_{aa} + \frac{p}{a}\Psi_a\biggl] -
\frac{\phi^2}{(3+2\omega)}\biggr[\Psi_{\phi\phi} + \frac{q}{\phi}\Psi_\phi
\biggl] - \frac{\phi^2}{2}\Psi_{\xi\xi} = V_\Psi(a)\Psi \quad ,
\end{equation}
where $V_\Psi(a) = 48a^4$.
\par
We will solve this equation for the cases $\xi = 0$ (one scalar field) and
$\xi \neq 0$ (two scalar fields).
\subsection{Solutions with one scalar field}
Discarding the field $\xi$, we have to solve the equation,
\begin{equation}
\frac{a^2}{12}\biggr[\Psi_{aa} + \frac{p}{a}\Psi_a\biggl] -
\frac{\phi^2}{(3+2\omega)}\biggr[\Psi_{\phi\phi} + \frac{q}{\phi}\Psi_\phi
\biggl] = V_\Psi(a)\Psi \quad .
\end{equation}
Supposing the separability of this equation, we can write
$\Psi(a,\phi) = \alpha(a)\beta(\phi)$
leading to two ordinary differential equations for $\alpha$ and
$\beta$:
\begin{eqnarray}
\alpha_{aa} + \frac{p}{a} &=& V_\alpha(a)\alpha \quad , \\
\beta_{\phi\phi} + \frac{q}{\phi}\beta_\phi &=& V_\beta(\phi)\beta
\quad ,
\end{eqnarray}
where
\begin{equation}
V_\alpha = 12\biggr(48a^2 - \frac{k}{a^2}\biggl)
\quad , \quad V_\beta(\phi) = - (3 + 2\omega)\frac{k}{\phi^2} \quad ,
\end{equation}
$k$ being an integration constant.
The solutions for $\alpha$ and $\beta$ are,
\begin{eqnarray}
\alpha_k(a) &=& a^{(1-p)/2}\biggr[A_\alpha I_n(12a^2) + B_\alpha
K_n(12a^2)\biggl]
\quad , \\
\beta_k(\phi) &=& A_\beta\phi^{(1-m-q)/2} +
B_\beta\frac{\phi^{(1+m-q)/2}}{m}
\quad ,
\end{eqnarray}
with $n = \frac{\sqrt{(p-1)^2 - 48k}}{4}$ and
$m = \sqrt{(q - 1)^2 - 4(3 + 2\omega)k}$.
The function $\alpha$ does not exhibit an oscillatory behaviour
unless $n \in {\bf I}$. For this case, $\alpha$ oscillates for
small values of $a$, increasing or decreasing for large values of $a$,
suggesting
that a classical phase may occur for small values of $a$ only. The
function $\beta$ has an oscillatory behaviour,
for all $\phi$, if $m \in {\bf I}$. In figure $1$ we show the
behaviour of the real and imaginary parts of $\alpha$ for $n \in$ {\bf I}.
The complete solution of the Wheeler-DeWitt equation is
\begin{equation}
\Psi(a,\phi) = \int{A(k)\alpha_k(a)\beta_k(\phi)dk} \quad .
\end{equation}

\begin{figure}
\includegraphics[trim=0 170 0 175, scale=0.95, clip]{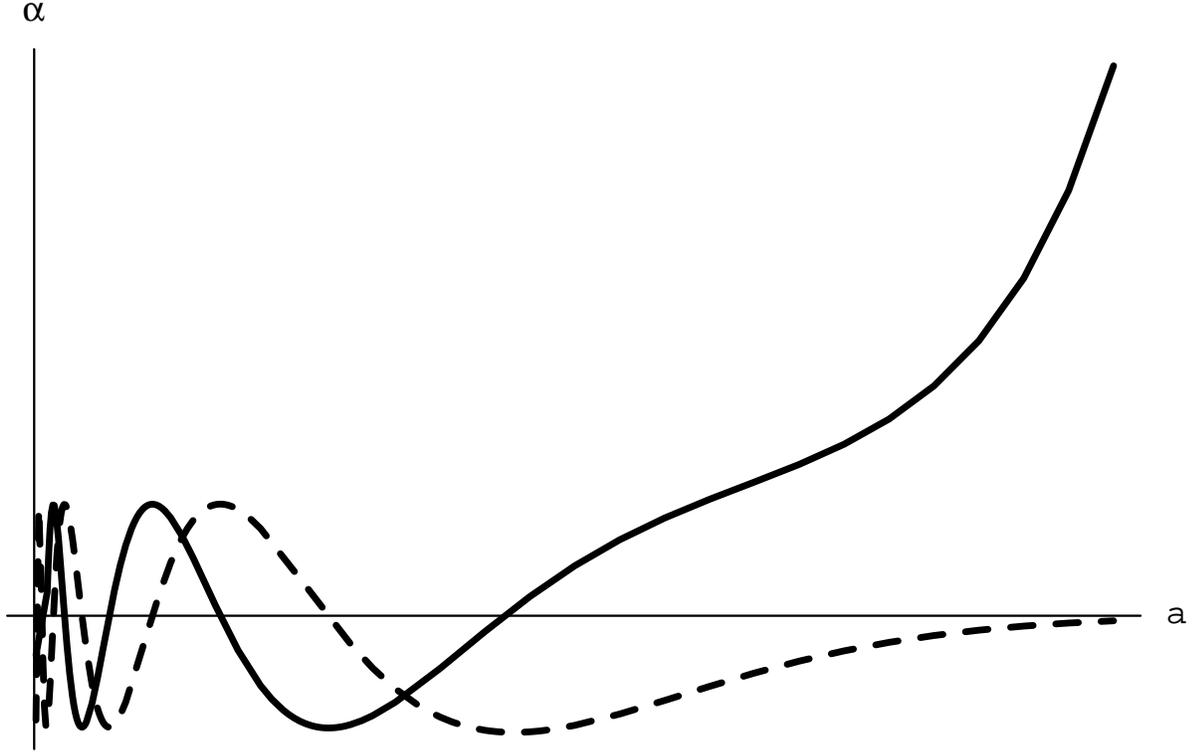}
\caption{Behaviour of $\alpha(a)$ for $n \in {\bf I}$ for the one scalar
field
case,
with $p = 1$, $k = 1$, $A_\alpha = 1$ and $B_\alpha = 0$.
The dashed and continous lines represent the imaginary 
and real parts of $\alpha$
respectively.}
\end{figure}

\subsection{Solutions with two scalar fields}
For the case where both scalar fields are non null, the
Wheeler-DeWitt equation in the minisuperspace reads,
\begin{equation}
\label{ww2}
\frac{a^2}{12}\biggr[\Psi_{aa} + \frac{p}{a}\Psi_a\biggl] -
\frac{\phi^2}{(3+2\omega)}\biggr[\Psi_{\phi\phi} + \frac{q}{\phi}\Psi_\phi
\biggl] - \biggr(\frac{\phi^2}{2}\biggl)\Psi_{\xi\xi} = V_\Psi(a)\Psi 
\quad .
\end{equation}
We use again the separation of variables method writing
$\Psi(a,\phi,\xi) = \Xi(a,\phi)\lambda(\xi)$.
Equation (\ref{ww2}) separates in two:
\begin{eqnarray}
\lambda_{\xi\xi} &=& - 8k_1\lambda \quad , \\
\frac{a^2}{12}\biggr[\Xi_{aa} + \frac{p}{a}\Xi_a\biggl] -
\frac{\phi^2}{(3+2\omega)}\biggr[\Xi_{\phi\phi} + \frac{q}{\phi}\Xi_\phi
\biggl] &=& V_\Xi(a,\phi)\Xi \quad ,
\end{eqnarray}
with $V_\Xi(a,\phi) = 48a^4 - 4k_1\phi^2$, $k_1$ being an integration
constant.
Writting $\Xi(a,\phi) = \alpha(a)\beta(\phi)$, we obtain two ordinary
equations:
\begin{eqnarray}
\alpha_{aa} + \frac{p}{\alpha}\alpha_a &=& V_\alpha(a)\alpha \quad , \\
\beta_{\phi\phi} + \frac{q}{\phi} &=& V_\beta(\phi)\beta \quad ,
\end{eqnarray}
with
\begin{eqnarray}
V_\alpha(a) &=& 12\biggr(48a^2 - \frac{k_2}{a^2}\biggl) \quad , \\
V_\beta(\phi) &=& (3 + 2\omega)\biggr(4k_1 - \frac{k_2}{\phi^2}\biggl)
\quad .
\end{eqnarray}
The solutions for $\alpha$, $\beta$ and $\lambda$ are
\begin{eqnarray}
\alpha(a) &=& a^{(1-p)/2}\biggr[A_\alpha I_n(12a^2) +
B_\alpha K_n(12\alpha^2)\biggl] \quad ,\nonumber \\ 
n &=& \frac{\sqrt{(p-1)^2 - 48k_2}}{4} \quad ; \\
\beta(\phi) &=& \phi^{1-q)/2}\bigg[A_\beta 
I_m\biggr(2\sqrt{(3+2\omega)k_1}\phi\biggl) +
B_\beta K_m\biggr(2\sqrt{3+2\omega k_1}\phi\biggl)\biggl] \quad
,\nonumber\\
m &=& \frac{\sqrt{(q-1)^2 - 4(3+2\omega)k_2}}{2} \quad ; \\
\lambda(\xi) &=& A_\lambda e^{i\sqrt{8k_1}\xi} +
B_\lambda e^{-i\sqrt{8k_1}\xi} \quad .
\end{eqnarray}
The coefficients $A$'s and $B$'s are constants. The general solution of
the Wheeler-DeWitt equation is
\begin{equation}
\label{fluminense}
\Psi(a,\phi,\xi) = \int{A(k_1,k_2)\alpha_{k_2}(a)\beta_{k_1,k_2}(\phi)
\lambda_{k_1}(\xi)dk_1dk_2} \quad .
\end{equation}
In general, $\alpha$ is an exponentially growing or
decreasing function of $a$. If the order of the
modified Bessel functions is imaginary, $\alpha$ may exhibit
an oscillatory behaviour. However, for these cases, $\alpha$
oscillates for small values of $a$, increasing or decreasing
for large values of $a$, suggesting again that a classical phase
may occur only for small $a$. This behaviour is displayed
in figure $2$.

\begin{figure}
\includegraphics[trim=0 170 0 175, scale=0.95, clip]{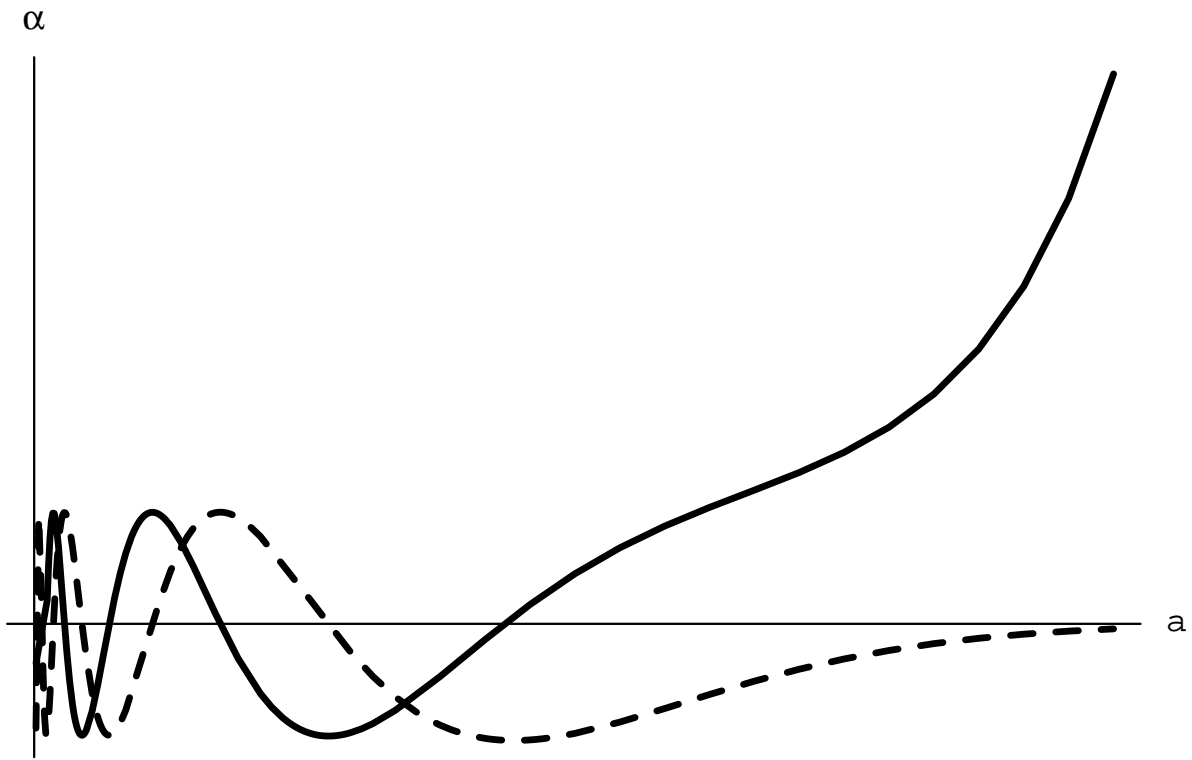}
\caption{Behaviour of $\alpha(a)$ for the two scalar fields case
for $p = 1$, $k_2 = 1$,
$A_\alpha = 1$ and $B_\alpha = 0$. The real part is
represented by the continous line while the imaginary part
is represented by the dashed line.}
\end{figure}

\section{The WKB Approximation}

One way to try to obtain the transition to the classical regime from
the quantum solutions is to employ the WKB approximation, like
in usual quantum mechanics. This is achieved by rewriting the
wave function as,
\begin{equation}
\Psi = \exp({\frac{i}{\hbar}S}) \quad ,
\end{equation}
substituting it into the Wheeler-DeWitt equation, and performing
an expansion in orders of $\hbar$ in $S$,
\begin{equation}
\label{wkbex}
S = S_0 + \hbar S_1 + \hbar^2S_2 + ...
\end{equation}
The classical solution must be recovered by constructing a wave packet
from $S_0$:
\begin{equation}
\Psi = \int{A(k_0)\exp{(\frac{i}{\hbar}S_0)dk_0}} \quad ,
\end{equation}
where $k_0$ is an integration constant.
As in the preceding sections, we will analyse the WKB approximation
separately for the cases with one and two scalar fields, respectively.

\subsection{WKB approximation with one scalar field}

In this case, we have $S = S(a,\phi)$, and the WKB expansion in
the minisuperspace Wheeler-DeWitt equation, leads to the
following equations connecting $S_0$ and $S_1$:
\begin{eqnarray}
\frac{a^2}{12}\biggr(\frac{\partial S_0}{\partial a}\biggl)^2 -
\frac{\phi^2}{3 + 2\omega}\biggr(\frac{\partial
S_0}{\partial\phi}\biggl)^2 + V_\Psi(a)
&=& 0 \quad ;\\
\frac{a^2}{12}\biggr[i\biggr(\frac{\partial^2S_0}{\partial a^2}\biggl)
-2\biggr(\frac{\partial S_0}{\partial a}\biggl)\biggr(\frac{\partial S_1}{
\partial a}\biggl) + \frac{ip}{a}\biggr(\frac{\partial S_0}{\partial
a}\biggl)\biggl]
- & & \nonumber \\
\frac{\phi^2}{3 +
2\omega}\biggr[i\biggr(\frac{\partial^2S_0}{\partial\phi^2}
\biggl) - 2\biggr(\frac{\partial
S_0}{\partial\phi}\biggl)\biggr(\frac{\partial S_1}{\partial\phi}\biggl)
+ \frac{iq}{\phi}\biggr(\frac{\partial S_0}{\partial\phi}\biggl)
\biggl] &=& 0 \quad .
\end{eqnarray}
First we get a solution for $S_0$. It can be obtained by taking,
\begin{equation}
S_0(a,\phi) = S_0(a) + S_0(\phi) \quad ,
\end{equation}
leading to two differential equations:
\begin{eqnarray}
\biggr(\frac{{\rm d} S_0(a)}{{\rm d} a}\biggl)^2 &=&
12\biggr(\frac{k_0}{a^2}
- 48a^2\biggl) \quad , \\
\biggr(\frac{{\rm d} S_0(\phi)}{{\rm d}\phi}\biggl)^2 &=& (3 + 2\omega)
\frac{k_0}{\phi^2} \quad ,
\end{eqnarray}
where $k_0$ is a separation constant.
These equations admit the following analytic solutions:
\begin{eqnarray}
\label{bri}
S_0(a) &=& \pm\bigg[\sqrt{3(k_0 - 48a^4)} - \sqrt{3k_0}
\mbox{arctanh}\biggr(\sqrt{\frac{k_0 - 48a^4}{k_0}}\biggl)\biggl] + A_0
\quad ,\\
\label{lul}
S_0(\phi) &=& \pm\sqrt{(3 + 2\omega)k_0}\ln\phi + B_0 \quad ,
\end{eqnarray}
where $A_0$ and $B_0$ are integration constants.
We follow the same procedure in order to obtain a solution for
$S_1(a,\phi)$,
considering first $S_1(a,\phi) = S_1(a) + S_1(\phi)$. We get
the solutions,
\begin{eqnarray}
S_1(a) &=& \pm\frac{k_1}{2}\sqrt{\frac{3}{k_0}}
\biggr[\mbox{arctanh}\biggr(\sqrt{\frac{k_0 - 48a^4}{k_0}}\biggl) +
i\frac{p - 1}{2}\ln a + \frac{i}{4}\ln(48a^4 - k_0)
+ A_1\biggl] \, ,\\
S_1(\phi) &=& \pm\biggr[i\frac{q - 1}{2} - \frac{k_1}{2}\sqrt{\frac{3 +
2\omega}{k_0}}\ln\phi\biggl] + B_1 \, ,
\end{eqnarray}
where $A_1$ and $B_1$ are integration constants.
From the solution for $S_0(a)$, we can easily see that only for $k_0 > 0$ 
we can obtain an oscillatory behaviour of the wavefunction for small
values of $a$, while for $k_0 < 0$ the wavefunction has an exponential
behaviour for any value of $a$. Similarly, if $(3 + 2\omega)k_0 > 0$,
then $\exp{[\frac{i}{\hbar}S_0(\phi)]}$ is oscillatory for any value of
$\phi$, otherwise it has an exponential behaviour. Hence,
for $k_0 > 0$ and $\omega > - \frac{3}{2}$,
$\exp{\frac{i}{\hbar}S_0(a,\phi)}$
oscillates for small values of $a$ and any value of $\phi$. 

We can construct a wavepacket from the above solutions through the
expression,
\begin{equation}
\label{wp}
\Psi(a,\phi) = \int{A(k_0)\exp{[\frac{i}{\hbar}S_0(k_0,a,\phi)]dk_0}}
\quad .
\end{equation}
where the function $A(k_0)$ is a sharply peaked gaussian centered in $\bar
k_0$, with width $\sigma$. Examining Eq. (\ref{bri}), we can see that $S_0
(a)$ 
becomes very large when $a$ becomes very small. Hence, in the integral
(\ref{wp}), constructive interference happens only if
\begin{equation}
\frac{\partial S_0(a,\phi)}{\partial k_0} = 0 \quad ,
\end{equation}
which implies a relation between $k_0, a$ and $\phi$, $k_0 = k_0(a,\phi)$.
The wave function turns out to be:
\begin{equation}
\label{wp2}
\Psi(a,\phi) =
A[k_0(a,\phi)]\exp\{\frac{i}{\hbar}S_0[k_0(a,\phi),a,\phi]\}
\quad .
\end{equation}
As the gaussian is sharply peaked at $k_0(a,\phi)=\bar k_0$, then we
obtain that
the wave function (\ref{wp2}) is sharply peaked at $k_0(a,\phi)=\bar k_0$.
It can be verified that this relation is exactly the classical relation
(\ref{implicit1}) with $\bar k_0$ playing the role of the integration
constant
$A$. 

\subsection{WKB approximation with two coupled scalar fields}
We follow the same procedure as before, writing the wave function
$\Psi$ in terms of $S(a,\phi,\xi)$, and performing an expansion in
orders of $\hbar$. The final equations for $S_0$ and $S_1$ are:
\begin{eqnarray}
\frac{a^2}{12}\biggr(\frac{\partial S_0}{\partial a}\biggl)^2 -
\frac{\phi^2}{3 + 2\omega}\biggr(\frac{\partial
S_0}{\partial\phi}\biggl)^2
 - \frac{\phi^2}{2}\biggr(\frac{\partial S_0}{\partial\xi}\biggl) +
V_\Psi(a) &=& 0 \quad ;\\
\frac{a^2}{12}\biggr[i\biggr(\frac{\partial^2S_0}{\partial a^2}\biggl)
-2\biggr(\frac{\partial S_0}{\partial a}\biggl)\biggr(\frac{\partial S_1}{
\partial a}\biggl) + \frac{ip}{a}\biggr(\frac{\partial S_0}{\partial
a}\biggl]
- & & \nonumber \\
\frac{\phi^2}{3 +
2\omega}\biggr[i\biggr(\frac{\partial^2S_0}{\partial\phi^2}
\biggl) - 2\biggr(\frac{\partial
S_0}{\partial\phi}\biggl)\biggr(\frac{\partial S_1}{\partial\phi}\biggl)
+ \frac{iq}{\phi}\biggr(\frac{\partial S_0}{\partial\phi}\biggl)
\biggl] + & & \nonumber \\
\frac{\phi^2}{2}\biggr[i\biggr(\frac{\partial^2S_0}{\partial\xi^2}\biggl)
- 2\biggr(\frac{\partial S_0}{\partial\xi}\biggl)\biggr(\frac{\partial
S_1}{\partial\xi}\biggl)\biggl] &=& 0 \quad .
\end{eqnarray}
Imposing again the ansatz $S_0(a,\phi,\xi) = S_0(a) + S_0(\phi)
+ S_0(\xi)$, we obtain the following equations:
\begin{eqnarray}
\biggr(\frac{\partial S_0(a)}{\partial a}\biggl)^2 &=&
12\biggr(\frac{K_0}{a^2}
- 48a^2\biggl) \quad , \\
\biggr(\frac{\partial S_0(\phi)}{\partial\phi}\biggl)^2 &=& (3 + 2\omega)
\biggr(\frac{K_0}{\phi^2} - k_0\biggl) \quad ,\\
\biggr(\frac{\partial S_0(\xi)}{\partial\xi}\biggl) &=& 2k_0 \quad ,
\end{eqnarray}
where $K_0$ and $k_0$ are separation constants.
The solutions are:
\begin{eqnarray}
S_0(a) &=& \pm\bigg[\sqrt{3(K_0 - 48a^4)} - \sqrt{3K_0}
\mbox{arctanh}\biggr(\sqrt{\frac{K_0 - 48a^4}{K_0}}\biggl)\biggl] + A_0
\quad , \nonumber \\
S_0(\phi) &=& \pm\bigg[\sqrt{(3+2\omega)(K_0 - k_0\phi^2)} -
\sqrt{(3+2\omega)K_0}
\mbox{arctanh}\biggr(\sqrt{\frac{K_0 - k_0\phi^2}{K_0}}\biggl)\biggl] +
B_0
\quad , \nonumber \\
S_0(\xi) &=& \pm\sqrt{2k_0}\xi + C_0 \nonumber,
\end{eqnarray}
where $A_0$, $B_0$ and $C_0$ are integration constants. As in the one
scalar
field case, we can 
find solutions for $S_1$ but they are not important for the construction
of the
wave packet in our approximation.
The solutions $S_0$ will be enough to recover the classical trajectory.
First we note that $K_0 > 0$ leads to a oscillatory behaviour for
$\exp{[\frac{1}{\hbar}S_0(a)]}$. On the other hand, if $(3 + 2\omega)K_0 >
0$, keeping $K_0 > 0$,
then $\exp{[\frac{i}{\hbar}S_0(\phi)]}$ is oscillatory for any value of
$\phi$ when $k_0 < 0$, or only for small values of $\phi$ when $k_0 > 0$.
If $(3 + 2\omega)K_0 < 0$, then $\exp{[\frac{i}{\hbar}S_0(\phi)]}$ has
an exponential behaviour for any value of $\phi$ when $k_0 < 0$ or
for small values of $\phi$ when $k_0 > 0$.
\par
We consider now the superposition given by
\begin{equation}
\label{cccpc}
\Psi(a,\phi,\xi) =
\int{\int{A(k_0,K_0)\exp{\frac{i}{\hbar}S_0(a,\phi,\xi,k_0,K_0)}
dk_0}dK_0},
\end{equation}
where $A(k_0,K_0)$ is a bidimensional gaussian function, centered on
$\bar k_0 > 0$ and $\bar K_0 > 0$ with width $\sigma_1$ and $\sigma_2$,
respectivelly. As before, $S_0 (a)$ becomes very large for small $a$.
Hence,
we have to guarantee constructive interference by the condition,
\begin{equation}
\label{che}
\bigg(\frac{\partial S_0(a,\phi,\xi)}{\partial k_0}\vert_{k_0=\bar k_0}
\biggl)^2 + \biggr(\frac{\partial S_0(a,\phi,\xi)}{\partial
K_0}\vert_{K_0=
\bar K_0}\biggl)^2 = 0 \quad .
\end{equation}
The implicit relations coming from (\ref{che}) are the same as
the classical relations (\ref{che1},\ref{che2},\ref{che3}). The classical
limit is again recovered only for small $a$.

\section{The Perspective of the Causal Interpretation}

In this section, we will apply the rules of the causal interpretation to
the
wave functions we have obtained in section $3$. We first summarize these
rules for the case of homogeneous minisuperspace models.
In the case of homogeneous models, the
supermomentum constraint ${\cal H}^i$ is identically zero, and the shift
function $N_i$ can be set to zero without loosing
any of the Einstein's equations. The hamiltonian is
reduced to general minisuperspace form:
\begin{equation} 
\label{homham}
H_{GR} = N(t) {\cal H}(p^{\alpha}(t), q_{\alpha}(t)),
\end{equation}
where $p^{\alpha}(t)$ and $q_{\alpha}(t)$ represent the homogeneous 
degrees of freedom coming from $\Pi ^{ij}(x,t)$ and $h_{ij}(x,t)$.
The minisuperspace Wheeler-De Witt equation is:
\begin{equation} 
\label{bsc}
{\cal H}({\hat{p}}^{\alpha}(t), {\hat{q}}_{\alpha}(t)) \Psi (q) = 0.
\end{equation}
Writing $\Psi = R \exp (iS/\hbar)$, and substituting it into (\ref{bsc}),
we obtain the following equation:
\begin{equation}
\label{hoqg}
\frac{1}{2}f_{\alpha\beta}(q_{\mu})\frac{\partial S}{\partial q_{\alpha}}
\frac{\partial S}{\partial q_{\beta}}+ U(q_{\mu}) +
Q(q_{\mu}) = 0,
\end{equation}
where
\begin{equation}
\label{hqgqp}
Q(q_{\mu}) = -\frac{1}{R} f_{\alpha\beta}\frac{\partial ^2 R}
{\partial q_{\alpha} \partial q_{\beta}},
\end{equation}
and $f_{\alpha\beta}(q_{\mu})$ and $U(q_{\mu})$ are the minisuperspace
particularizations of the DeWitt metric $G_{ijkl}$ \cite{dew} and of the
scalar curvature density $-h^{1/2}R^{(3)}(h_{ij})$ of the spacelike
hypersurfaces,
respectively.
The causal interpretation, applied to quantum cosmology, states that
the trajectories $q_{\alpha}(t)$ are real, independently of any
observations.
Eq. (\ref{hoqg}) is the Hamilton-Jacobi equation for them, which
is the classical one
ammended with a quantum potential term (\ref{hqgqp}), responsible
for the quantum effects. This suggests
to define:
\begin{equation}
\label{h}
p^{\alpha} = \frac{\partial S}{\partial q_{\alpha}} ,
\end{equation}
where the momenta are related to the velocities in the usual way:
\begin{equation}
\label{h2}
p^{\alpha} = f^{\alpha\beta}\frac{1}{N}\frac{\partial q_{\beta}}{\partial
t} \; .
\end{equation}
To obtain the quantum trajectories we have to solve the following
system of first order differential equations:
\begin{equation}
\label{h3}
\frac{\partial S(q_{\alpha})}{\partial q_{\alpha}} =
f^{\alpha\beta}\frac{1}{N}\frac{\partial q_{\beta}}{\partial t} \; .
\end{equation}
Eqs. (\ref{h3}) are invariant under time reparametrization. Hence,
even at the quantum level, different choices of $N(t)$ yield the same 
spacetime geometry for a given non-classical solution $q_{\alpha}(t)$.
There is no problem of time in the causal interpretation of minisuperspace
quantum cosmology.
Let us then apply this interpretation to our minisuperspace models
and
choose the gauge $N = 1$.

\subsection{One scalar field}

The general solution of the Wheeler-DeWitt equation is given by
\begin{equation}
\label{gen1}
\Psi(a,\phi) = \int{A(k)\alpha_k(a)\beta_k(\phi)dk} \quad .
\end{equation}
where
\begin{eqnarray}
\label{cccpa}
\alpha_k &=& a^{\frac{1-p}{2}}\biggr[A_\alpha I_n(12a^2) +
B_\alpha K_n(12a^2)\biggl] \quad ,\\
\label{cccpb}
\beta_k &=& A_\beta\phi^{\frac{1-m-q}{2}} +
\frac{B_\beta}{m}\phi^{\frac{1+m-q}{2}} \quad ,
\end{eqnarray}
with
\begin{equation}
n = \frac{\sqrt{(p-1)^2 - 48k}}{4} \quad 
\end{equation}
and
\begin{equation}
m = \sqrt{(q-1)^2 - 4(3+2\omega)k} \quad .
\end{equation}
The momenta are
\begin{eqnarray}
\pi_a &=& 24a\dot a \quad ,\\
\pi_\phi &=& - 2(3+2\omega)a^3\frac{\dot\phi}{\phi^2} \quad .
\end{eqnarray}
The causal interpretation states that the momenta are also given by
\begin{eqnarray}
\pi_a &=& \frac{\partial S(a,\phi)}{\partial a} \quad , \\
\pi_\phi &=& \frac{\partial S(a,\phi)}{\partial\phi} \quad ,
\end{eqnarray}
where $S(a,\phi)$ is the total phase of the wave function $\Psi$. Hence,
the Bohmian trajectories will be solutions of the following system of
equations:
\begin{eqnarray}
\label{et1}
24a\dot a &=& \frac{\partial S(a,\phi)}{\partial a} \quad , \\
-2(3 + 2\omega)a^3\frac{\dot\phi}{\phi^2} &=& \frac{\partial
S(a,\phi)}{\partial\phi} \quad .
\label{et2}
\end{eqnarray}
The quantum potential for this problem can be calculated in the usual way.
We substitute $\Psi = Re^{iS}$ into the Wheeler-DeWitt
equation,
obtaining the Hamilton-Jacobi like equation with the extra quantum
potential
term $Q$:
\begin{equation}
-\frac{a^2}{12}\biggr(\frac{\partial S}{\partial a}\biggl)^2 +
\frac{\phi^2}{3+2\omega}\biggr(\frac{\partial S}{\partial\phi}\biggl)^2 -
48a^4 + Q = 0 \quad ,
\end{equation}
where
\begin{equation}
\label{pq}
Q = \frac{1}{R}\biggr[\frac{a^2}{12}\biggr(\frac{\partial^2R}{\partial
a^2}
+ \frac{p}{a}\frac{\partial R}{\partial a}\biggl) -
\frac{\phi^2}{3+2\omega}\biggr(\frac{\partial^2R}{\partial\phi^2} +
\frac{q}{\phi}\frac{\partial R}{\partial\phi}\biggl)\biggl] \quad .
\end{equation}
Let us apply this interpretation to the simplest case
$\Psi = \alpha_k(a)\beta_k(\phi)$. Then the wavefunction has the form,
\begin{equation}
\Psi = R_1(a)R_2(\phi)e^{i[S_1(a) + S_2(\phi)]} \quad ,
\end{equation}
since $S(a,\phi) = S_1(a) + S_2(\phi)$. This implies that (\ref{et1})
becomes independent of $\phi$. From Eq. (\ref{pq}), we see that
$Q(a,\phi) = Q_1(a) + Q_2(\phi)$.
To simplify
the calculations we set $A_\beta = B_\beta = 0$, and $p = q = 1$.
We will first calculate the dynamics of the scale factor when $a$ is
small,
in order to see if there are singularities. In this
approximation we can take just the first term of the series
represantion of $I_n(x)$ \cite{gra},
\begin{equation}
\label{bf1}
I_n(x) =
\sum_{l=0}^\infty\frac{1}{l!\Gamma(n+l+1)}\biggr(\frac{x}{2}\biggl)^{n+2l}
\quad .
\end{equation}
For $n$ real, the modified Bessel function $I_n(x)$ is real and the phase
of
$\alpha _k$ is zero. Hence, the Bohmian equation (\ref{et1}) yields that
$a$
is a constant. It is a nonsingular quantum solution but with little
physical
interest. Hence, in a first moment, we will restrict ourselves to the case
where $n$ is a pure imaginary number. Combinations of these two situations 
will be analyzed afterwards.

In the case where $n$ is pure imaginary, $\alpha_k$ can be written as
\begin{equation}
\alpha_k = c_0x^{i\nu} = c_0e^{i\nu\ln x} \quad , \quad n = i\nu = \pm
i\sqrt{3k}
\quad .
\end{equation}
The phase and the norm are,
\begin{eqnarray}
\begin{array}{l}
S_1(a) = \nu\ln x \quad , \\
R_1(a) = c_0 \quad .
\end{array}
\end{eqnarray}
Defining $x=12a^2$, Eq. (\ref{et1}) becomes 
\begin{equation}
\dot a = \frac{{\rm d} S}{{\rm d} x} = \frac{\nu}{x} \quad ,
\end{equation}
whose solution is
\begin{equation}
\label{a1}
a = (\frac{\nu t}{4})^{\frac{1}{3}} \quad .
\end{equation}
We can see that, in accordance to what was suggested in previous sections,
the
behaviour of the quantum trajectory is like the
classical one for small $a$, and the singularity will still be present.
Note that if $\nu$ is positive we have expansion, while if $\nu$ is
negative we have contraction.
For the scalar field, we have:
\begin{equation}
\label{f}
\beta _k = A_\beta \phi^{iu} = B_\alpha e^{iu\ln\phi} \quad ,
\end{equation}
where $u = - \sqrt{(3+2\omega)k}$. The phase and the norm are:
\begin{equation}
\label{p1}
\begin{array}{l}
S_2(\phi) = u\ln\phi \quad ,\\
R_2(\phi) = B_\alpha \quad .
\end{array}
\end{equation}
From Eq. (\ref{et2}) we have,
\begin{equation}
- 2\frac{3+2\omega}{\phi^2}a^3\dot\phi = \frac{\partial S}{\partial\phi}
= \frac{u}{\phi} \quad .
\end{equation}
Using Eq. (\ref{a1}) we get
\begin{equation}
\label{f1}
\phi = t^{\frac{2}{\sqrt{3(3 + 2\omega)}}} \quad ,
\end{equation}
which is also the classical behaviour for $A = 1$.
\par
It is not surprising that we have obtained the classical behaviour.
Since $R_1(a)$ and $R_2(a)$ are constants, the quantum potential is
zero, and there is no quantum effect. Note also that solutions
(\ref{a1}) and (\ref{f1}) satisfy the hamiltonian constraint with $V(a) =
48a^4$ neglected because $a$ is very small.
\par
For very large $a$, the Bessel function $I_n(z)$ can be approximated to
\cite{gra}
\begin{equation}
I_n(x) \sim c_1\frac{e^{12a^2}}{a} \quad ,
\end{equation}
where $c_1$ is a complex constant.
In this case, we have:
\begin{eqnarray}
S_1(a) &=& {\rm const.} \quad , \\
R_1(a) &=& |c_1|\frac{e^{12a^2}}{a} \quad .
\end{eqnarray}
The quantum trajectory is evidently 
\begin{equation}
\label{a2}
a = {\rm const.} \quad ,
\end{equation}
which is not
the classical one.
For the scalar field, Eqs. (\ref{et2},\ref{p1},\ref{a2}) now yields,
\begin{equation}
\phi = e^{\frac{1}{2}\sqrt{\frac{k}{3+2\omega}}t} \quad ,
\end{equation}
which is the classical behaviour for $\phi$ in this regime (note that as
$a =$ const., $\theta \propto t$ in equation (\ref{fi1})). The different
behaviours of the scale factor and the scalar field can be explained with
the quantum potential. For the scale factor, $Q_1$ will be given 
(see Eq. (\ref{pq}) with $p = 1$):
\begin{equation}
Q_1(a) = 48a^4 + \frac{1}{12}
\end{equation}
which is of the same size of the classical potential $V(a) = - 48a^4$, 
and hence responsible for this quantum behaviour. For the scalar field,
as $R_2(\phi)$ is constant, $Q_2(\phi) = 0$ and the scalar field 
continues to follow its classical trajectory. Note that the hamiltonian
constraint is also satisfied in this limit. Hence
we have again obtained the strange result where the classical limit
happens
only for small values of $a$.

\begin{figure}
\includegraphics[bb=100 80 530 600, scale=.75, angle=-90]{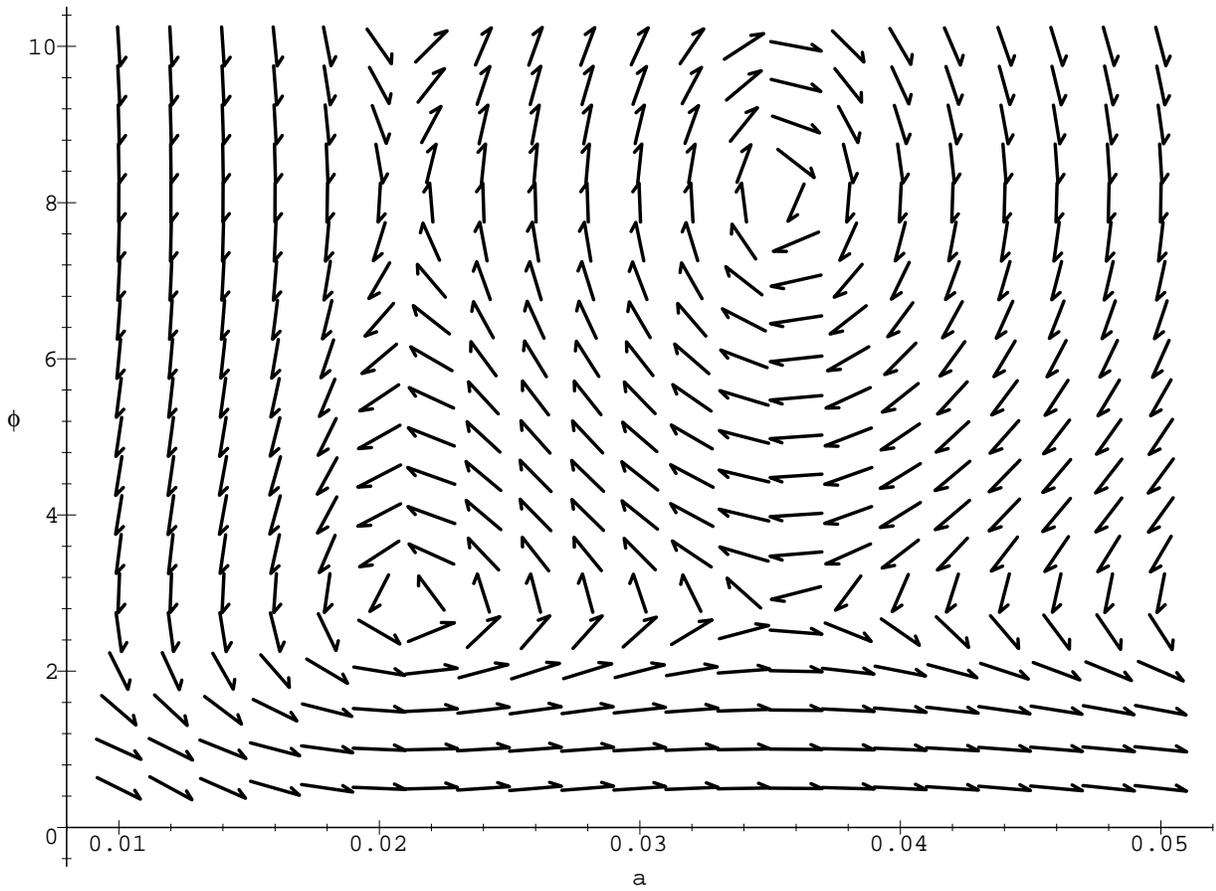}
\caption{Field plot of $a$ versus $\phi$ using the causal interpretation
for the superpositions of the wave functions in the one scalar field case,
in the region of small $a$.}
\end{figure}

\begin{figure}
\includegraphics[bb=100 80 560 600, scale=.75, angle=-90]{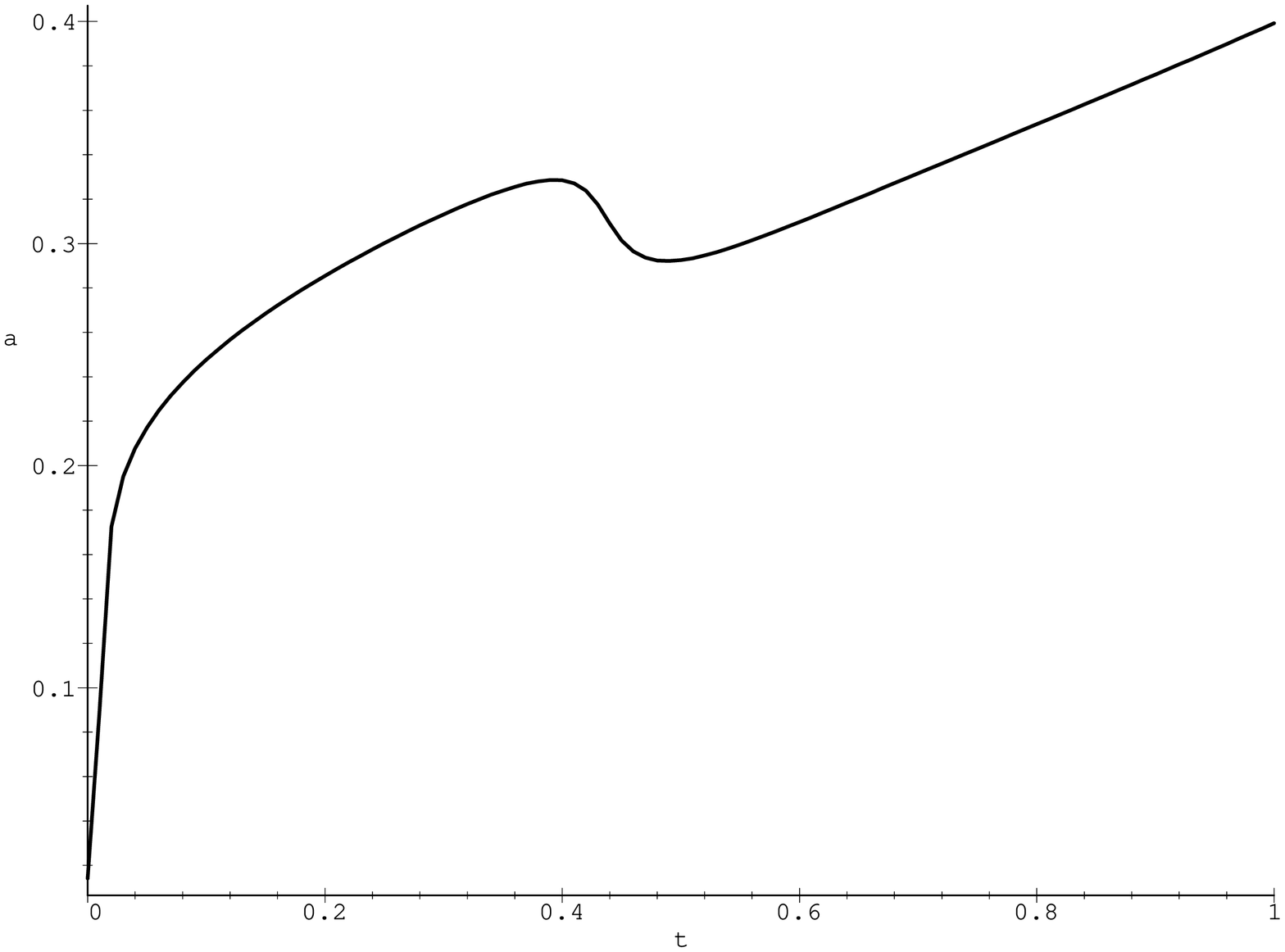}
\caption{Plot of a particular singular solution for $a(t)$, coming from
figure $3$, which begins with an inflationary phase.}
\end{figure}

\par

Let us now take some superpositions of the $\alpha_k(a)$ and
$\beta_k(\phi)$
given in Eqs. (\ref{cccpa},\ref{cccpb}). For definitiness, we will choose
$p = q = 1$, and $\omega = 0$. We will continue to take only pure
imaginary $n$'s. Combinations of real and pure imaginary $n$'s do not
change qualitatively the results. The wave function will be given by,
\begin{equation}
\Psi = \sum_{i=1}^3 I_{n_i}(12a^2)\biggr[A_i\phi^{-u_i} +
B_i\phi^{u_i}\biggl]
\end{equation}
where
\begin{equation}
\left.
\begin{array}{l}
k_1 = \frac{1}{3} \rightarrow n_1 = i \quad , \quad u_1 = i \quad ,\\
k_2 = k_3 = \frac{e^{2\pi i}}{3} \rightarrow n_2 = n_3 = - i \quad , \quad 
u_2 = u_3 = - i \quad , \\
A_1 = A_2 = 1 \quad , \quad A_3 = 0 \quad , \\
B_1 = B_2 = 0 \quad , \quad B_3 = 1 \quad .
\end{array}
\right\}
\end{equation}
Using Eq. (\ref{bf1}) we obtain, for small $a$:
\begin{equation}
\label{fo2}
\Psi = \biggr(\frac{12a^2}{\phi}\biggl)^ i + 
\frac{1}{(12a^2\phi)^ i} + \biggr(\frac{\phi}{12a^2}\biggl)^i \quad .
\end{equation}
Figure $3$ shows a field plot of $a$ versus $\phi$, for small $a$, for the
Bohmian equations (\ref{et1},\ref{et2}), with $S$ being the phase of the
wave function (\ref{fo2}). We can
see that there are periodic solutions with very small oscillations around
$a$. They are eternal quantum universes which never grows. The solutions
which grow beyond the validity of (\ref{fo2})
are singular and inflationary. This can be seen in figure $4$.
\par
This result suggests that the initial singularity can be avoided only if
we
superpose eigenfunctions of opposite frequencies. However, it seems to be
be difficult to obtain, in scalar field models, non-singular universes
with long expansion period \cite{bola2}.

\subsection{Two Scalar Fields}
In this case, we have studied the quantum trajectories driven by wave
functions
obtained from Eq. (\ref{fluminense}) for some particular $A(k_1,k_2)$.
\par i) We have fixed $p = q = 1$, $\omega = 0$, $k_1 = - \frac{1}{12}$,
$k \equiv k_2$, $A_\alpha = A_\beta = 0$, $B_\alpha = B_\beta = A_\lambda
=B_\lambda = 1$ and $A(k_1,k_2) = \frac{3}{2}\delta(k_1 + \frac{1}{12})
\tanh(\pi\sqrt{3k_2})$. Using a result of Ref. \cite{gra}, we obtain,
\begin{eqnarray}
\Psi(x,\phi,\xi) &=& \cosh\biggr(\sqrt{\frac{2}{3}\xi}\biggl)\int_0^\infty
\frac{3}{2}\tanh(\pi\sqrt{3k})K_{i\sqrt{3k}}(x)K_{i\sqrt{3k}}(i\phi)dk
\quad ,
\nonumber \\
&=&\cosh\biggr(\sqrt{\frac{2}{3}}\xi\biggl)\frac{\pi}{2}\sqrt{\frac{x\phi}{x^2
+ \phi^2}}
e^{-x}\exp\biggr\{i\biggr[\frac{\pi}{4} - \phi - {\rm
arctan}(\frac{\phi}{x})\biggl]\biggl\}
\quad .
\end{eqnarray}
The quantum trajectories can be calculated from the following equations
(in the gauge $N = 1$):
\begin{equation}
\begin{array}{l}
\pi_a = 24a\dot a = \frac{\partial S}{\partial a} =
\frac{24a\phi}{x^2 + \phi^2} \quad ,\\
\pi_\phi = - 6 \frac{a^3\dot\phi}{\pi^2} = \frac{\partial S}{\partial\phi}
=
\frac{x^2 + \phi^2 + x}{x^2 + \phi^2} \quad , \\
\pi_\xi = - 4\frac{a^3\dot\xi}{\phi^2} = \frac{\partial S}{\partial \xi}
=
0.
\end{array}
\end{equation}
The solutions are:
\begin{equation}
\label{s3}
\begin{array}{l}
a = \frac{1}{\sqrt{12}}\biggr[\ln\biggr(\frac{C}{\sqrt{1 +
4\eta ^2}}\biggl)\biggr]^{\frac{1}{2}} \quad , \\
\phi = - \frac{1}{2\eta}\ln\biggr(\frac{C}{\sqrt{1 + 4\eta ^2}}\biggl) =
- 6\frac{a^2}{\eta} \quad ,\\
\xi = {\rm const.} \quad ,
\end{array}
\end{equation}
where $\eta = \int\frac{dt}{a}$ is the conformal time and $C$ is an
integration
constant. 
For small $a$, when $\eta$ approaches $\pm \frac{\sqrt{c^2 -1}}{2}$, these
functions
tends to:
\begin{equation}
\begin{array}{l}
a(t) \propto t^{\frac{1}{3}} \quad , \\
\phi(t) \propto t^{\frac{2}{3}} \propto a^2 \quad ,\\
\xi = {\rm const.} \quad .
\end{array}
\end{equation}
which is exactly the classical behaviour for $\omega = 0$. When $a$ is not
small, the trajectories are not classical (compare (\ref{s3}) with
Eqs. (\ref{flu1},\ref{flu2},\ref{flu3})). This can be seen by inspecting
the quantum potential. For
two scalar
fields it is given by
\begin{equation}
Q = \frac{1}{R}\biggr[\frac{a^2}{12}\biggr(\frac{\partial^2 R}{\partial
a^2}
+ \frac{p}{a}\frac{\partial R}{\partial a}\biggl) -
\frac{\phi^2}{3+2\omega}\biggr(\frac{\partial^2R}{\partial\phi^2} +
\frac{q}{\phi}\frac{\partial R}{\partial \phi}\biggl)
- \frac{\phi^2}{2}\frac{\partial^2R}{\partial\xi^2}\biggl] \quad .
\end{equation}
For our particular problem, we obtain
\begin{equation}
Q = \frac{1}{3}\frac{(x^4 - 2\phi^2x + \phi^4)}{x^2 + \phi^2} \quad .
\end{equation}
For small $a$ we have $Q \propto a^2$ (remember that $\phi \propto a^2$
in this limit).
In this domain, the kinetic terms dominate:
\begin{eqnarray}
K_a &=& \frac{a^2\pi_a^2}{12} = \frac{a^2}{12}\biggr(\frac{\partial
S}{\partial
a}\biggl)^2 = \frac{1}{3}\frac{x^2\phi^2}{(x^2 + \phi^2)^2} \propto
{\rm const.} \quad , \\
K_\phi &=& - \frac{\phi^2\pi_\phi^2}{3} = -
\frac{\phi ^2}{3}\biggr(\frac{\partial
S}{\partial\phi}\biggl)^2 = - \frac{\phi^2}{3}\biggr(\frac{x^2 + \phi^2 +
x}{
x^2 +\phi^2}\biggl)^2 \propto {\rm const.} \quad .
\end{eqnarray}

Hence the quantum potential becomes negligible when compared with the
classical kinetic terms.
For $a$ not small, for instance, when $n \rightarrow 0$ yielding $a
\rightarrow a_{max}$ and $\phi \rightarrow \infty$, the quantum potential
diverges,
\begin{equation}
Q \propto \phi^2 \quad ,
\end{equation}
while the classical potential and kinetic terms behave like
\begin{eqnarray}
K_a &\propto& \frac{1}{\phi^2} \quad , \\
K_\phi &\propto& \phi^2 \quad ,\\
V_{cl} &\propto& a^4 \quad .
\end{eqnarray}
Hence, together with $K_\phi$, the quantum potential becomes the more
important term.
This behaviour of the quantum potential explains why the trajectories
are classical for small $a$ and quantum otherwise.
\par
ii) Let us now take $p = q = 1$, $\omega = 0$, $k_1 = \frac{i}{6}$,
$k \equiv k_2$, $A_\alpha = A_\beta = B_\lambda = 0$, $B_\alpha = B_\beta
=
A_\lambda
= 1$ and $A(k_1,k_2) = \delta(k_1 - \frac{i}{6})\sinh(\pi\sqrt{3k_2})
K_{i\sqrt{3k_2}}(\sqrt{2}e^{i\frac{\pi}{4}})$. 
Then we obtain the following wave function (see Ref. \cite{gra}):
\begin{equation}
\Psi = \frac{\pi^2}{4}\exp\biggr[-\frac{x}{2}(\phi + \frac{1}{\phi})
-\sqrt{\frac{2}{3}}\xi\biggl]\exp\biggr[i\biggr(\sqrt{\frac{2}{3}}\xi -
\frac{\phi}{x}\biggl)\biggl] \quad .
\end{equation}
The equations of motion are:
\begin{equation}
\begin{array}{l}
\pi_a = 24a\dot a = \frac{\partial S}{\partial a} = 24\frac{a\phi}{x^2}
\quad ,\\
\pi_\phi = - 6\frac{a^3\dot\phi}{\phi^2} = \frac{\partial
S}{\partial\phi}
= - \frac{1}{x} \quad ,\\
\pi_\xi = - 4\frac{a^3\dot\xi}{\phi^2} = \frac{\partial S}{\partial\xi}
= \sqrt\frac{2}{3} \quad .
\end{array}
\end{equation}
The solutions are,
\begin{equation}
\begin{array}{l}
a = t^\frac{1}{3} \quad , \\
\phi = C_0a^2 = c_0t^\frac{2}{3} \quad ,\\
\xi = - |c_1|t^\frac{4}{3} + C_2 \quad .
\end{array}
\end{equation}
Note again that these solutons approach the classical one when $a$ 
is small but are completely different when $a$ is large. Once
again,
this can be explained by the behaviour of the quantum potential when
compared with the kinetic and classical potential terms. The quantum
potential
is given by
\begin{equation}
\label{pq3}
Q = 48a^4 - \frac{\phi^2}{3} \quad .
\end{equation}
The kinetic and classical potential terms are given by,
\begin{equation}
\label{s4}
\begin{array}{l}
K_a = \frac{\phi^2}{3x^2} \quad ,\\
K_\phi = - \frac{\phi^2}{3x^2} \quad ,\\
K_\xi = \frac{\phi^2}{3} \quad ,\\
V_a = - 48a^4 \quad .
\end{array}
\end{equation}
For small $a$, $Q$, $V_a$ and $K_\xi$ goes to zero while $K_a$ and
$K_\phi$
are constant.
For large $a$, $Q$, $V_\phi$ and $K_\xi$ become comparable and large,
while
$K_a$ and $K_\phi$ continue to be constant. Hence, in this situation, the
quantum
potential becomes important, driving the quantum behaviour of the Bohmian
trajectories.
Note that the sum of (\ref{pq3}) with (\ref{s4}) gives zero because the
Bohmian
trajectories must satisfy the hamiltonian constraint ammended with the
quantum potential term. We have also calculated the Bohmian trajectories
for
other exact wave solutions of the Wheeler-DeWitt equation. All of them
present
the same behaviour.
\par
We conclude this section by stating that in quantum cosmology it is not
necessary
that the classical behaviour appears when $a$ is large, while quantum
behaviour is present when $a$ is small. It can indeed be the reverse. This 
was already pointed out in \cite{glik} and we presented specific examples
illustrating this fact. It should also be commented that the result of
this seciton using the causal
interpretation are in qualitative agreement with the results of the
previous
section.

\section{Conclusion}

In this paper, we have studied classical and quantum minisuperspace models
containing one and two scalar fields. We have shown that all classical
solutions are singular. After quantizing the models, we have obtained
the general solution of the Wheeler-De Witt equation. Usually, the
solutions
are oscillatory when the scale factor is small and not oscillatory
when the scale factor becomes larger. This suggests that non-classical
behaviour may occur when the sacale factor is large. We studied
gaussian superpositions of WKB wave functions to investigate if they
correspond to quasi-classical states, as suggested in Ref. \cite{kief}.
We have shown that indeed these wave functions
are peaked around the classical trajectories in configuration space,
but only for small $a$.

After, we applied the causal interpretation of quantum mechanics to
these models. In this interpretation, it is possible to calculate
quantum trajectories, independently of any observations. We have shown
that the trajectories calculated following this interpretation usually
present the classical behaviour when the scale factor is small
and non-classical behaviour when $a$ is large, as suspected. This means 
that these quantum trajectories still presents an initial singularity.
We have also seen that if we superpose negative with positive frequency
solutions, then we can find trajectories which are no more classical for
small $a$.
We can have eternal periodic quantum universes with very small
oscilations.
These universes, however, never scape the Planck length. There 
are also singular solutions with a short period of inflation which grow
forever. We could not find any non-singular solution which grows to the 
size of our universe, with a classical limit for large $a$.

The fact that quantum behaviour happens when $a$ is large is not
surprising. It was already obtained in Ref. \cite{bola2} and suggested
to exist when the scale factor grows like $t^{1/3}$ in Ref. \cite{glik},
which is our case.
Hence, in quantum
cosmology it is not necessarily true that large scale factors implies
classical behaviour. For the scalar field models we have analyzed in this
paper, the reverse seems to be more usual. It means that it is possible to
have in our universe
some degrees of freedom which still behave quantum mechanically
in spite of it being very big. This gives us some hope of being
possible to detect or experience quantum cosmological effects in the
real universe we live in, bringing quantum cosmology to the realm
of testable physical theories. The problem should be to find which
degrees of freedom can possess this property. To do this, we should
improve this minisuperspace model with the accretion of small
inhomogeneous
perturbations, which contain an infinity number of degrees
of freedom, and see what happens with the new inhomogeneous degrees of
freedom. We will get closer to the real universe but we will have
to face new technical and interpretational problems. This will be the
subject
of our future investigations.
\newline
{\bf Acknowledgements} We thank CNPq (Brazil) and CAPES (Brazil)
for financial support.

\end{document}